\documentclass[twocolumn,showpacs,preprintnumbers,amsmath,amssymb,floatfix]{revtex4}

\usepackage{graphicx}
\usepackage{epsfig} 
\usepackage{dcolumn}
\usepackage{bm}
\usepackage{epstopdf}
\usepackage{multirow}
\usepackage{booktabs}

\begin{document}

\preprint{APS/123-QED}

\title{The $d \: ^3 \Pi$ state of LiRb}

\author{I. C. Stevenson$^{1,}$\footnotemark[1], D. Blasing$^{2,}$\footnote[1]{These authors contributed equally to this work}, A. Altaf, Y. P. Chen$^{2,1,3}$ and D. S. Elliott$^{1,2,3}$}
\affiliation{%
  $^1$School of Electrical and Computer Engineering, $^2$Department of Physics and Astronomy and
  $^3$Purdue Quantum Center, \\ Purdue University, West Lafayette, IN  47907
}

\date{\today}

\begin{abstract}
We report our spectroscopic studies of the $d \ ^3\Pi$ state of ultra-cold $^7$Li$^{85}$Rb using resonantly-enhanced multi-photon ionization and depletion spectroscopy with bound-to-bound transitions originating from the metastable $a \ ^3\Sigma^+$ state.  We evaluate the potential of this state for use as the intermediate state in a STIRAP transfer scheme from triplet Feshbach LiRb molecules to the $X \ ^1\Sigma^+$ ground state, and find that the lowest several vibrational levels possess the requisite overlap with initial and final states, as well as convenient energies.  Using depletion measurements, we measured the well depth and spin-orbit splitting. We suggest possible pathways for short-range photoassociation using deeply-bound vibrational levels of this electronic state.
\end{abstract}

\maketitle

\section{Introduction}
Ultra-cold molecules can play important roles in studies of many-body quantum physics~\cite{pupillo2008cold}, quantum logic operations~\cite{demille2002quantum}, and ultra-cold chemistry~\cite{ni2010dipolar}.  In our recent studies of LiRb, motivated largely by its large permanent dipole moment~\cite{aymar2005calculation}, we have explored the generation of these molecules in a dual species MOT~\cite{dutta2014formation,dutta2014photoassociation}.  In particular, we have found that the rate of generation of stable singlet ground state molecules and first excited triple state molecules through photoassociation, followed by spontaneous emission decay, can be very large~\cite{v0paper,Adeel,lorenz2014formation}.  There have been very few experimental studies of triplet states in LiRb~\cite{Adeel}, in part because they are difficult to access in thermally-distributed systems.  Triplet states of bi-alkali molecules are important to study for two reasons: first, Feshbach molecules, which are triplet in nature, provide an important association gateway for the formation of stable molecules~\cite{marzok2009feshbach}; also, photoassociation (PA) of trapped colliding atoms is often strongest for triplet scattering states.  Mixed singlet - triplet states are usually required to transfer these molecules to deeply bound singlet states.
\begin{figure}[t!]
	\includegraphics[width=8.6cm]{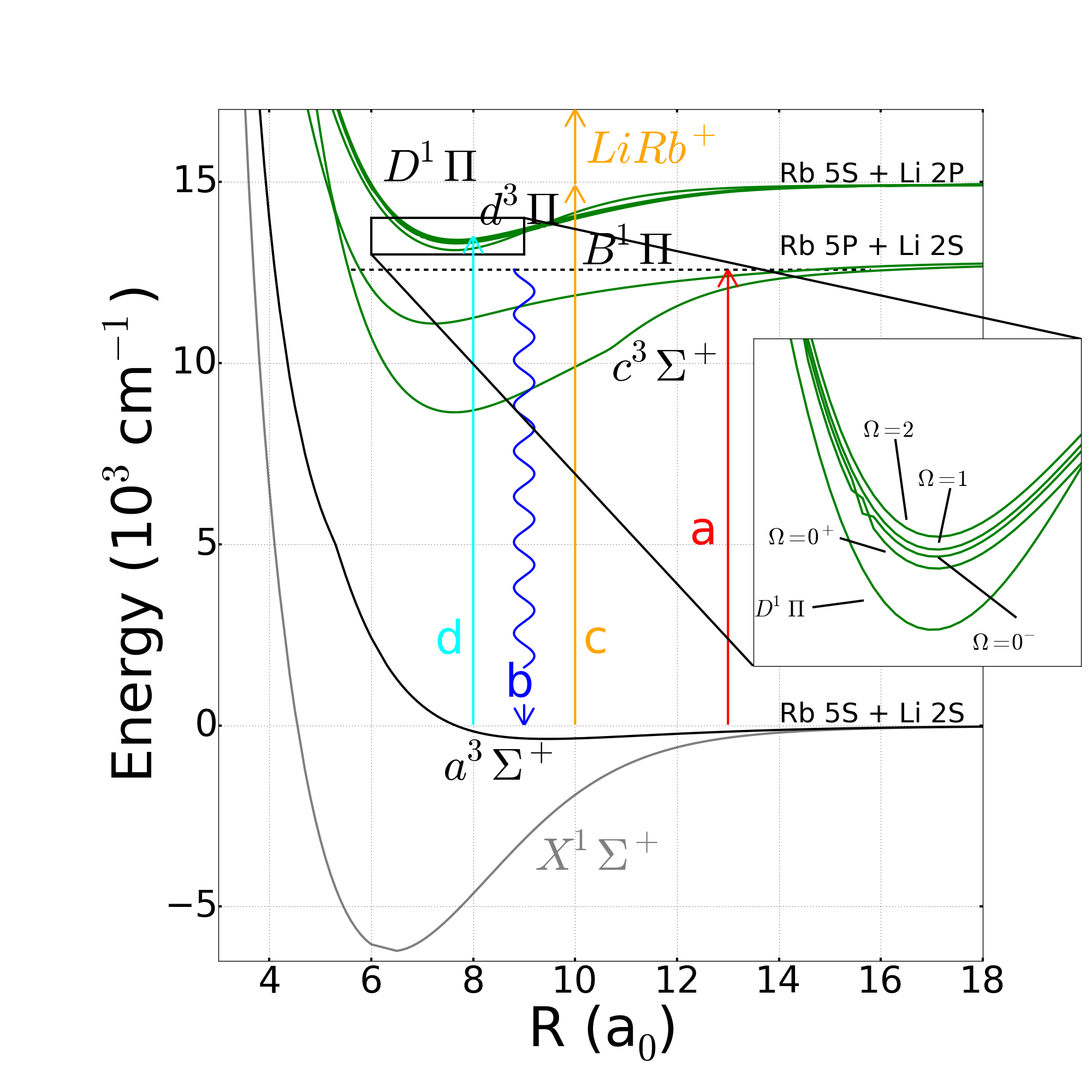}\\
	\caption{(Color on-line) Energy level diagram of the LiRb molecule, showing relevant PECs from Ref.~\protect\cite{Korek}.  Vertical lines show the various optical transitions, including {\bf (a)} photoassociation of atoms to molecular states below the  D$_1$ asymptote; {\bf (b)} spontaneous decay of excited state molecules leading to the $a \: ^3 \Sigma ^+$ state; {\bf (c)} RE2PI to ionize LiRb molecules, ($\nu_{c}$ used later in this paper is the frequency of this laser source); and {\bf (d)} state-selective excitation of the $a \: ^3 \Sigma ^+$ state for depletion of the RE2PI signal (with laser frequency $\nu_{d}$).  The black dashed line represents our PA states.  The inset shows an expanded view of the different $d \ ^3\Pi$ spin-orbit split states as well as the perturbing neighbor $D \ ^1\Pi$.}
	\label{fig:PEC}
\end{figure}
\begin{figure*} [t!]
	\includegraphics[width=\textwidth]{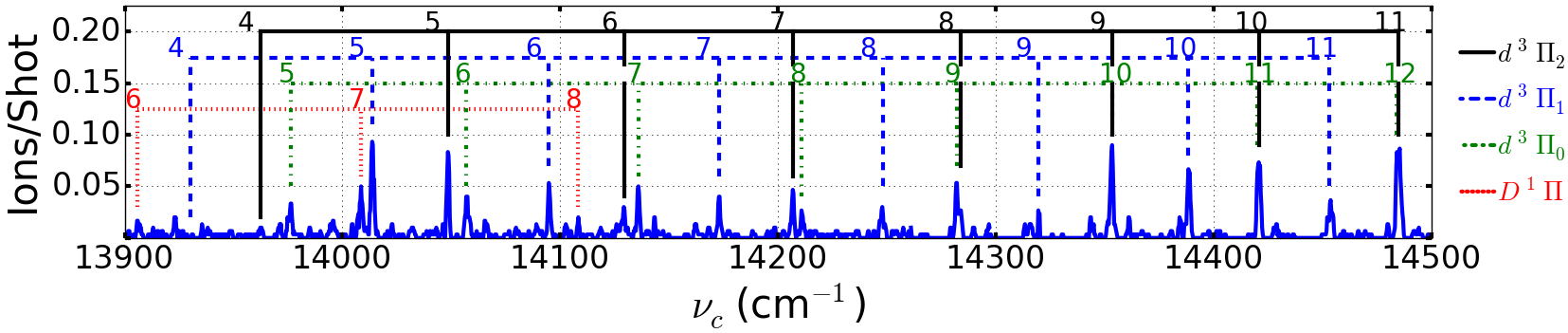}\\
	\caption{(Color on-line) Subsection of the RE2PI spectra.  The PA laser is tuned to the $2(0^-) \ v=-11 \ J=1$ resonance, from which spontaneous decay is primarily to the $a \ ^3\Sigma^+ \ v^{\prime \prime}=11$ state.  Most of these lines are $d \ ^3\Pi_{\Omega} \ v^{\prime} \leftarrow a \ ^3\Sigma^+ \ v^{\prime \prime}=11$ transitions, where $v^\prime$ is labeled on individual lines.  From top to bottom: black solid lines label transitions to $\Omega=2$, blue dashed lines label transitions to $\Omega=1$, green dot-dashed lines label transitions to $\Omega=0$.  Also shown (red dotted lines) are three $ D \ ^1\Pi \ v^\prime \leftarrow a \ ^3\Sigma^+ \ v^{\prime \prime}=11$ transitions.}
	\label{fig:v11progression}
\end{figure*}

We show an abbreviated set of potential energy curves (PEC), as calculated in Ref.~\cite{Korek}, in Fig.~\ref{fig:PEC}.  The d$^3 \Pi$ - D$^1 \Pi$ complex in LiRb, asymptotic to the Li 2p $^2P_{3/2, 1/2}$ + Rb 5s $^2S_{1/2}$ free atom state, has several features that can promote its utility in stimulated-Raman-adiabatic-passage (STIRAP) and photoassociation.  First, the \textit{ab inito} calculations of Ref.~\cite{Korek} predict mixing between low vibrational levels of the $d \ ^3\Pi_1$ and the D$^1 \Pi$ states.  Second, both legs of a STIRAP transfer process from loosely bound triplet-character Feshbach molecules to the rovibronic ground state can be driven with commercially-available diode lasers.  And third, similar deeply bound $^3 \Pi$ resonances have been successfully used for short-range PA in RbCs ~\cite{RbCs1,RbCs2,RbCs3}.  While an interesting discovery on its own, spontaneous decay of these states after PA can populate the $a \ ^3\Sigma^+ \ v^{\prime \prime}=0$ state; one RbCs team~\cite{RbCs2} found spontaneous decay of these states even populated the $X \ ^1\Sigma^+ \ v^{\prime \prime}=0$ state.    

In the present work, we study the $d \ ^3\Pi_{\Omega}$ states of LiRb, from the asymptote to the most bound vibrational level. We have found signatures of state mixing between low-lying vibrational levels of the d$^3 \Pi_1$ and D$^1 \Pi$ levels.  We have determined the term energies of the different spin-orbit components of the $d \ ^3\Pi$ state, as well as their vibrational energies.  We have also observed alternation of the intensities of the rotational lines, a possible indication of a $p$-wave resonance in the scattering state, and determined the rotational constants of the lowest vibrational levels.

\section{Experiment}
We have previously described the details of our experimental apparatus~\cite{DuttaALEC14b}, and provide here only a brief summary.  We trap $\sim 5 \times 10^7$ Li atoms and $\sim 2 \times 10^8$ Rb atoms in a dual species magneto-optical trap (MOT), $\lesssim$1 mK in temperature and 1 mm in diameter~\cite{Adeel}.  Our Rb MOT is a spatial dark spot MOT~\cite{Ketterle}.  We photoassociate Li and Rb atoms to form LiRb molecules using either a 300 mW cw Ti:Sapphire laser or a 150 mW cw external cavity diode laser.   After spontaneous decay to a distribution of vibrational levels of the $a \ ^3\Sigma^+$ state, we use two-color  resonantly-enhanced-multi-photon-ionization (RE2PI) to ionize the molecules.  The lasers that we use to drive the RE2PI process are a Nd:YAG-pumped, pulsed dye laser (PDL) for the first photon, tunable in the wavelength range between 667 nm - 750 nm (14950 - 13300 cm$^{-1}$ frequency range), and part of the 532 nm pump laser for the second photon.  The repetition rate of this system is 10 Hz, and it delivers $\sim$1.5 mJ/pulse of dye energy in a 4 mm diameter beam and a larger $\sim$2 mJ/pulse 532 nm beam to the MOT region.  When the frequency of the dye laser is resonant with a transition from an initial state (populated through spontaneous decay from the PA state) to an intermediate bound state (in this experiment the $d \ ^3\Pi$ state), absorption of a dye photon and a 532 nm photon ionizes the molecule.  We detect the molecular LiRb$^+$ ions using a time-of-flight spectrometer and a microchannel plate detector.  In this paper we will adopt the following notation: $v^{\prime \prime}$ and $J^{\prime \prime}$ denote the vibrational and rotational levels of the $a \: ^3 \Sigma ^+$ and $X \ ^1\Sigma^+$states, $v$ and $J$ (without a prime) denote the vibrational and rotational levels of the PA resonances (and for these vibrational numbers, we count down from the asymptote using negative integers), and $v^{\prime}$ and $J^{\prime}$ denote vibrational and rotational labeling of other excited electronic states.
\begin{figure*} [t]
	\includegraphics[width=\textwidth]{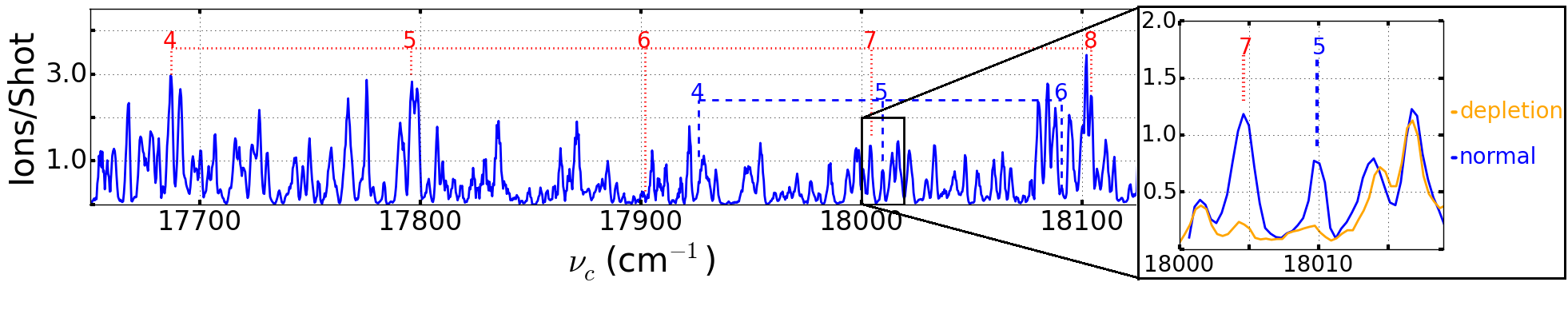}\\
	\caption{(Color on-line) Subsection of REMPI data with the PA laser tuned to the $4(1) \ v=-16 \ J=1$ resonance~\cite{v0paper}, while scanning the REMPI laser frequency, $\nu_c$.  Transitions are labeled as $d \ ^3\Pi_{1} \ v^{\prime} \leftarrow X \ ^1\Sigma^+ \ v^{\prime \prime}=10$ (blue dashed) and $D \ ^1\Pi \ v^{\prime} \leftarrow X \ ^1\Sigma^+ \ v^{\prime \prime}=10$ (red dotted).  The inset shows confirmation of the assignments with depletion spectroscopy.  The orange curve in the inset is the REMPI data retaken in the presence of a depletion laser tuned to the $A \ ^1\Sigma^+ \ v^{\prime}=25 \ J^{\prime}=1 \leftarrow X \ ^1\Sigma^+ \ v^{\prime \prime}=10 \ J^{\prime \prime}=0$~\cite{A1Sigma+} transition; the reduction in peak height confirms the assignments.  Because the depletion laser reduces the population available for REMPI in both peaks, they must have the same initial state, from which we conclude that we can access triplet REMPI resonances from singlet states.}
	\label{fig:v10progression}
\end{figure*}

We used two techniques, RE2PI and depletion spectroscopy, to measure the $d \ ^3\Pi$ bound states.  In RE2PI spectroscopy, we tune the PA laser to either the $v=-11 \ J=1$ or $v=-8 \ J=1$ lines of the $2(0^-)$ long range state, from which spontaneous decay leads primarily to the vibrational levels of the $a \ ^3\Sigma^+$ state~\cite{Adeel}.  We count the number of ions detected over the course of 100 laser pulses, and tune the laser frequency $\nu_c$ of the PDL in 0.35 cm$^{-1}$ increments.  We record the number of ions detected, normalized by the number of laser pulses, as a function of the PDL frequency $\nu_c$.

In order to reach the full range of vibrational levels of the $d \ ^3\Pi$ states, we used two different laser dyes in the PDL.  An LDS 698 dye covered the 13950 - 14950 cm$^{-1}$ range, and an LDS 750 dye covered from 13300 to 13950 cm$^{-1}$.  These dyes are difficult to work with because of short lifetimes and low power output.  The LDS 750 dye in particular was very troublesome: it has a lifetime $\le$8 hours, produces low power ($\le$0.5 mJ/pulse for much of its range) and because it has a very broad pulse width (i.e. lots of spontaneous emission) the baseline noise of our RE2PI spectra is enhanced over what we have observed with other dyes.  

The $2(0^-) \ v=-11 \ J=1$ PA line at $\nu_a = 12516.89$ cm$^{-1}$ is relatively weak, but it decays almost exclusively to a single vibrational level ($v^{\prime \prime}=11$) of the $a \ ^3\Sigma^+$ state.  This facilitates straight-forward identification of the vibrational levels of the intermediate state. Unfortunately, several vibrational levels of the $d \ ^3\Pi$ state do not appear in this spectrum, presumably due to poor Franck-Condon overlap with the $v^{\prime \prime}=11$ state.  This problem was even more evident when using the LDS 750 dye in the PDL.  For this reason, we collected several RE2PI spectra using the stronger $2(0^-) \ v=-8 \ J=1$ PA resonance $\nu_a = 12557.60$ cm$^{-1}$. This line decays to a wider spread of vibrational levels, giving more complete coverage of the vibrational lines of the $d \ ^3\Pi$ states, but also making our analysis more difficult, due to the increased congestion of the spectra and frequent overlap between individual lines.

To explore the deeply-bound levels of the $d \ ^3\Pi$ states, we used a depletion spectroscopy technique.  In these measurements, we used the 150 mW ECDL tuned to the $2(0^-) \ v=-5 \ J=1$ PA resonance at $\nu_a = 12575.05$ cm$^{-1}$~\cite{Adeel}.  Spontaneous decay of this state populates the $a \ ^3\Sigma^+ \ v^{\prime \prime}=13$ state.   We tune the PDL laser frequency to the $(3) \ ^3\Pi_{0} \ v^{\prime}=6 \leftarrow a \ ^3\Sigma^+ \ v^{\prime \prime}=13$ one-color resonantly-enhanced two-photon ionization (REMPI) transition at 17736.6 cm$^{-1}$~\cite{Adeel}.  We then tune the frequency of the Ti:Sapphire laser into resonance  with bound-to-bound transitions from the $a \ ^3\Sigma^+ \ v^{\prime \prime}=13$ state to ro-vibrational levels in the $d \ ^3\Pi$ state.  Exciting these transitions depletes the population of the $a \ ^3\Sigma^+ \ v^{\prime \prime}=13$ state, causing the REMPI signal to decrease.

\section{RE2PI Measurements}

We show an example of a RE2PI spectrum in Fig.~\ref{fig:v11progression}.  Transitions observed in this spectrum are $ d \ ^3\Pi_{\Omega} \ v^{\prime} \leftarrow a \ ^3\Sigma^+ \ v^{\prime \prime}=11$.  We have marked the transitions to the $\Omega=2$ progression with black solid lines, $\Omega=1$ with blue dashed lines, and $\Omega=0$ with green dot-dashed lines.  $\Omega$ is the total electronic angular momentum, orbital $L$ + spin $S$, projected onto the internuclear axis.  The numerical label for each peak is the vibrational number $v^{\prime}$ of the  $d \ ^3\Pi_{\Omega}$ state.  We have also marked three lines in this spectrum corresponding to transitions to the $D \ ^1\Pi$ state with red dotted lines.

From the spectrum of these $d \ ^3\Pi$ states, we observe the typical hierarchy of line spacings: the vibrational splitting is large (on the order of 100 cm$^{-1}$ for low vibrational quantum number $v^\prime$, and decreasing with increasing $v^\prime$), and the spin-orbit splitting between different $\Omega$ states is smaller (on the order of 30 cm$^{-1}$).  The rotational splitting for low $J^\prime$ (on the order of 0.1 cm$^{-1}$) is too small to be resolved in these RE2PI spectra since the spectral resolution of the PDL is $\sim$0.5 cm$^{-1}$.  

The appearance of transitions belonging to vibrational levels of the $D \ ^1\Pi$ electronic state in Fig.~\ref{fig:v11progression} is evidence of mixing between the $D \ ^1\Pi$ and the $d \ ^3\Pi_{1}$ potentials near an avoided crossing between the two states.  The energy of these $D \ ^1\Pi$ states is known from Refs.~\cite{Ivanova11,JohnThesis}.  State mixing gives these states partial character of each electronic state, which in this case manifests itself through strong transitions from a triplet state (i.e. $a \ ^3\Sigma^+ \ v^{\prime \prime}=11$) to singlet states ($D \ ^1\Pi \ v^{\prime}$ = 6, 7, and 8).  This state mixing also adds $D ^1\Pi$  character to the $d \ ^3\Pi_{1}$ states, so one should expect the nearby $d \ ^3\Pi_{1}$ states to appear in the singlet spectra.  This expectation is borne out in the spectrum shown in Fig.~\ref{fig:v10progression}.  This spectrum is a REMPI scan generated in our system after photoassociating ultracold LiRb molecules through a 2(1) - 4(1) mixed state at $\nu_a = 12574.85$ cm$^{-1}$~\cite{v0paper}, which spontaneously decays to vibrational levels of the $X \ ^1\Sigma^+$ ground electronic state.  The spectrum in Fig.~\ref{fig:v10progression} primarily shows transitions to low-lying $D \ ^1\Pi$ vibrational levels from $X \ ^1\Sigma^+ \ v^{\prime \prime}=10$.  We also observe in this spectrum $d \ ^3\Pi_{1} \ v^{\prime}=4$, 5, and 6 $\leftarrow X \ ^1\Sigma^+ \ v^{\prime \prime}=10$ transitions.  We chose $X \ ^1\Sigma^+ \ v^{\prime \prime}=10$ because it is strongly populated by spontaneous decay of the 2(1) - 4(1) PA resonance and transitions to deeply bound $D \ ^1\Pi$ vibrational levels are clearly identified. We can estimate the degree of mixing based on the relative strength of the different REMPI peaks.  The $D \ ^1\Pi \ v^\prime = 7 \leftarrow X \ ^1\Sigma^+ \ v^{\prime \prime}=10$ transition is twice as strong as the $d \ ^3\Pi_1 \ v^\prime = 5 \leftarrow X \ ^1\Sigma^+ \ v^{\prime \prime}=10$ transition so there is twice as much singlet character to $D \ ^1\Pi \ v^\prime = 7$ as $d \ ^3\Pi_1 \ v^\prime = 5$.  Following the same procedure that we have used in the past~\cite{v0paper}, we can estimate the interaction strength to be about 7 cm$^{-1}$.  Interestingly, this rough estimate is consistent with the following simple perturbative argument.  The spin-orbit interaction responsible for the state mixing can be estimated as about one half the spin orbit mixing in atomic rubidium~\cite{WeidemullerReview}, or about 120 cm$^{-1}$.  The Franck - Condon factor (FCF) between  $D \ ^1\Pi \ v^\prime = 7$ and $d \ ^3\Pi_1 \ v^\prime = 5$, as calculated by LEVEL 8.0~\cite{LEVEL} using the PEC from~\cite{Korek}, is about 0.08 and thus the strength of interaction between these states should be approximately 10 cm$^{-1}$.  We have applied this perturbative analysis to each of the vibrational levels of the $d \ ^3\Pi_{1} $ state (not including $v^{\prime}=5$), and find that each contains some small component of $D \ ^1\Pi$ perturber state, on the order of 10\% or smaller.  This is too small to be seen in the spectra of Fig.~\ref{fig:v11progression}, but could be sufficient to be useful in a Raman or STIRAP transfer of population to low lying levels of the electronic ground state in the future.

Many of the RE2PI spectra that we collected are less clear than that shown in Fig.~\ref{fig:v11progression}.  In particular, the peaks in the RE2PI spectra near the Rb 5S + Li 2P asymptote are strong, but line congestion becomes significant, and clear identification of the lines in this region becomes difficult.  These assignments could probably be improved using a spectroscopic technique that is capable of higher spectral resolution, such as photoassociative spectroscopy, but this was beyond the scope of the present work. Assigning peaks in the RE2PI spectra was equally difficult for deeply bound vibrational states (i.e. $v^{\prime}$ $\le$ 4).  In fact we were unable to observe a clear cut off in our RE2PI data corresponding to $v^{\prime}=0$.  We attribute this to difficulties with the LDS 750 dye, specifically the large spontaneous emission content in the pulse.  To rectify this problem, we turned to a second set of measurements, based upon depletion spectroscopy.  

\section{Depletion Measurements}

\begin{figure} [b]
	\includegraphics[width=8.6cm]{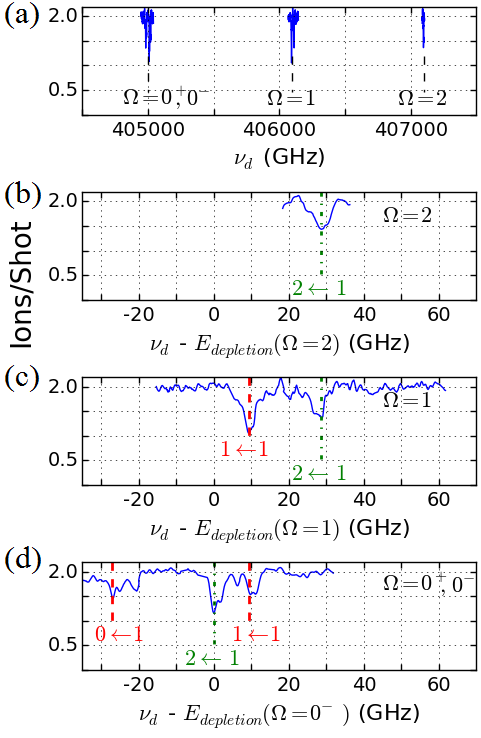}\\
	\caption{Depletion spectra of an $a \ ^3\Sigma^+ \ v^{\prime \prime}=13$ REMPI line using the $d \ ^3\Pi \ v^{\prime}=0$ state.  The PA laser is locked to the $2(0^-) \ v=-5, \ J=1$ line, the REMPI laser is tuned to the $(3) \ ^3\Pi_{0} \ v^{\prime}=6 \leftarrow v^{\prime \prime}=13$ transition at $\nu_c = 17736.6$ cm$^{-1}$.  Panel (a) shows a global view of our depletion data.  Panels (b) - (d) show the rotational structure of the depletion lines.  All repeatable transitions are labeled $J^{\prime} \leftarrow K^{\prime \prime}$.  For transitions $ J^\prime = 0$ or 1 $\leftarrow K^{\prime \prime} = 1$ (red lines), $J^{\prime \prime}=0$ or 1 are possible.  For transitions $J^\prime = 2 \leftarrow K^{\prime \prime} = 1$ (green lines), $J^{\prime \prime}$ can only be 1.  The abscissa of (b) - (d) is the depletion laser frequency offset by $E_{depletion} = T_{0, \ \Omega} - T_{13} - 2B_{13}$ where $T_{0, \ \Omega}$ is the rotationless energy of the $d \ ^3\Pi_{\Omega} \ v^{\prime}=0$ state, $T_{13}$ and $B_{13}$ are the binding energy and the rotational constant of the $a \ ^3\Sigma^+ \ v^{\prime \prime}=13$ state respectively.}
	\label{fig:depletion}
\end{figure}
\begin{table*}[t]
	\centering
	\begin{tabular}{ccccccccc}
		\hline \hline
		& \multicolumn{2}{c}{$d \ ^3\Pi_{0^+}$}& \multicolumn{2}{c}{$d \ ^3\Pi_{0^-}$} &\multicolumn{2}{c}{$d \ ^3\Pi_{1}$} &\multicolumn{2}{c}{$d \ ^3\Pi_{2}$} \\ 
		$v^{\prime}$ & $T_v$ ($cm^{-1}$) & $B_{v}$ ($cm^{-1}$) & $T_v$ ($cm^{-1}$) & $B_{v}$ ($cm^{-1}$) & $T_v$ ($cm^{-1}$) & $B_{v}$ ($cm^{-1}$)& $T_v$ ($cm^{-1}$) & $B_{v}$ ($cm^{-1}$)\\ \midrule[1.pt]
		0 & 13507.9 & 0.148 (4) & 13508.8 &  & 13544.7 & 0.153 (6)& 13577.6 & \\ 
		1 & 13606.8 & 0.146 (4) & 13607.6 &  &  &  & & \\ \hline \hline
	\end{tabular}
	\caption{Experimental assignments for $T_v$, the rotationless energy, and $B_v$, the rotational constant, of the vibrational levels of the $d \ ^3\Pi$ state based on our depletion data for $v^{\prime}=0$ and 1.  Uncertainty for all $T_v$ is 0.5 cm$^{-1}$, and uncertainty in the last digit in $B_v$ is given in parentheses.  Blank entries denote rotational constants or energies that we were not able to measure because of either the tuning range of our Ti:Sapphire laser or because of bound to bound selection rules limited by our PA state.}
	\label{tab:depletionAssignments}
\end{table*}

We used depletion spectroscopy to identify the lowest two vibrational levels of the $d \ ^3\Pi$ state. Compared to RE2PI, depletion spectra are sparse and have narrow peak widths, in this case $\sim$1 GHz, a typical linewidth for this type of measurement at this intensity~\cite{Deiglmayr}.  Additionally, depletion spectra allow us to extract some rotational constants of the lines, a very useful tool for comparing experiment to theory. 

\begin{figure} [t]
	\includegraphics[width=8.6cm]{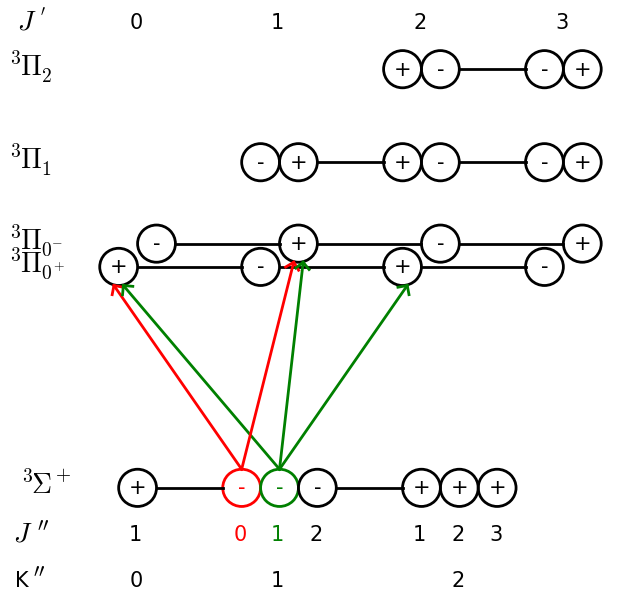}\\
	\caption{Rotational structure and parity of $^3\Sigma^+$ and $^3\Pi$ vibrational states, the two states in red and green highlight the ground states we populate at the beginning of the depletion process.  Transitions from the populated ground states to $^3\Pi_{0^{+}}$ and $^3\Pi_{0^{-}}$ are shown by arrows.  Transitions to the positive symmetry levels of $^3\Pi_{1}$ and $^3\Pi_{2}$ do occur, but are omitted for clarity.  The rotational splitting in $^3 \Sigma^+$ is determined by $E=B_vK^{\prime \prime}(K^{\prime \prime}+1)$ (which makes the $J$ manifold within each $K^{\prime \prime}$ state degenerate) and in $^3 \Pi$ by $E=B_v[J(J+1)-\Omega^2]$.  Adapted and modified from Ref.~\cite{Herzberg}}
	\label{fig:parity}
\end{figure}

We show depletion spectra for $v^{\prime}=0$ in Fig.~\ref{fig:depletion}.  To assign these data care must be taken with selection rules for radiative transitions in molecules.  The two that apply here are: $\Delta J= 0, \ \pm1$ and $- \leftrightarrow +$, that is positive symmetry states (with respect to coordinate inversion) must transition to negative symmetry states and vise versa.  The initial state in this depletion transition is a $^3\Sigma^+$ state, which is a strict Hund's case (b) state.  As such, its rotational energy is determined by quantum number $K$, which designates the total angular momentum of the molecule apart from spin, rather than the total angular momentum (including spin) quantum number $J$.  For this $a ^3\Sigma^+$state, the electronic spin is $S$=1, and levels with $J=K+1, K$ and $K-1$, are nearly degenerate for $K\ge 1$.  Additionally $K$ determines the symmetry of the state.  This is summarized in Fig.~\ref{fig:parity}, adapted and modified from Ref.~\cite{Herzberg}.  

The logic that leads to our assignments in Fig.~\ref{fig:depletion} goes as follows.  We start with the depletion data on transitions to $\Omega=1$ shown in Fig.~\ref{fig:depletion}(c), which is extensive enough to show that we do not populate $J^{\prime \prime}=2$ of the $a \ ^3\Sigma^+$ state; that is, we see no depletion signal corresponding to a transition to $J^{\prime} = 3$. Since we do see transitions to $J^{\prime} = 1$ and $2$, we know that we populate some mix of $J^{\prime \prime}=0$ and $1$ belonging to either $K^{\prime \prime}=0$, 1, or 2.  $K^{\prime \prime}=0$ and $K^{\prime \prime}=2$ have the same symmetry and some of the same rotational numbers; this implies that any PA state that could decay to $K^{\prime \prime}=0$ or 2 could decay to the other as well.  Because the energies of $K^{\prime \prime}=0$ and $K^{\prime \prime}=2$ differ from one another, but we see no additional structure in the depletion spectra, we infer that the transitions seen in Fig.~\ref{fig:depletion}(c) originate from a single state, that is $K^{\prime \prime}=1$.  The spacing between the two peaks in this spectrum should be $4B_0$, where  $B_0$ is the rotational constant of the $d \ ^3\Pi \ v^\prime = 0$ state, allowing us to determine $B_0 = 4.59$ GHz.  This rotational constant agrees with the prediction from LEVEL 8.0 with PECs from Ref.~\cite{Korek}, which confirms our assignments and guides our interpretation of the $\Omega=0$ data shown in Fig.~\ref{fig:depletion}(d), in which we see three transitions.  The first two are spaced by $6B_0$ (27.5 GHz), implying that these peaks are transitions to the positive symmetry levels of the $\Omega=0^+$ electronic state, $J^{\prime}=0 \leftarrow K^{\prime \prime}=1$ and $J^{\prime}=2 \leftarrow K^{\prime \prime}=1$.  The absence of a peak for $J^{\prime}=1$ is consistent with the selection rule $- \not\leftrightarrow -$.  The remaining transition in $\Omega=0$ must be the only allowed transition to $\Omega=0^-$, that is $J^{\prime}=1 \leftarrow K^{\prime \prime}=1$.  There is only one transition in $\Omega=2$, which is trivial to identify as $J^{\prime}=2 \leftarrow K^{\prime \prime}=1$.  This picture is consistent with what we know about our PA state.  That is, we PA into the $J=1$ level of a $2(0^-)$ state, which has positive symmetry, and the allowed decay paths are to $K^{\prime \prime}=1$ (with $J^{\prime \prime}=0$, $1$, and $2$, each with negative symmetry, of which we populate $0$ and $1$).  Of note, we chose to PA to $J=1$ because it is the strongest PA resonance in the $2(0^-) \ v=-5$ progression.  Our depletion data confirms that the parity of this PA state is even, which is suggestive of a scattering state $p$-wave shape resonance~\cite{d3PiPA}.  Similar shape resonances have been seen in other bi-alkali's~\cite{NaCs,Deiglmayr}.

We used these data to determine the spin-orbit splitting between the different $\Omega$ progressions deep in the $d \ ^3\Pi$ well, and followed these progressions back to the asymptote in our RE2PI data.  Most importantly, these data provide accurate locations of $v^{\prime}=0$ and $1$, $J^{\prime}$ for future work on short range PA~\cite{d3PiPA}.

The depletion data is limited by the shot noise in our ion counting.  For each data point we integrate 200 shots and usually count $\sim$350 ions.  To get 2$\sigma$ resolution the smallest depletion feature that we can be confident in has to be a 10\% decrease.  Here our on-resonance depletion signal drops by around 30\% which is 6$\sigma$ and statistically significant.  Unfortunately, our Ti:Sapphire struggles to tune to wavelengths much shorter than 740 nm, which limited our depletion spectra to the lowest two vibrational levels only.  Despite these short comings, depletion spectroscopy gives us accurate identifications of these bottom two vibrational levels of the $d \ ^3\Pi$ electronic state and unambiguously determines the spin orbit splitting.
\begin{table*}[t]
	\centering
	\begin{tabular}{ccccccc}
		\hline \hline
		& \multicolumn{2}{c}{$d \ ^3\Pi_0$} &\multicolumn{2}{c}{$d \ ^3\Pi_1$} &\multicolumn{2}{c}{$d \ ^3\Pi_2$} \\ 
		$v^{\prime}$ & $T_v$ ($cm^{-1}$) & $\Delta$E ($cm^{-1}$) & $T_v$ ($cm^{-1}$) & $\Delta$E ($cm^{-1}$) & $T_v$ ($cm^{-1}$) & $\Delta$E ($cm^{-1}$)\\ \midrule[1.5pt]
		0 & 13508.2 & 98.8 & 13545.2 & 99.1 & 13578.1 & 99.3\\ 
		1 & 13607.0 & 94.9 & 13644.3 & 95.4 & 13677.4 & 92.0\\ 
		2 & 13701.9 & 94.8 & 13739.7 & 92.0 & 13769.4 & 96.0\\ 
		3 & 13796.7 & 87.3 & 13831.7 & 89.6 & 13865.4 & 88.1\\ 
		4 & 13883.9 & 83.8 & 13921.3 & 83.5 & 13953.5 & 86.1\\ 
		5 & 13967.7 & 80.5 & 14004.8 & 81.2 & 14039.6 & 80.9\\ 
		6 & 14048.2 & 79.0 & 14086.0 & 78.1 & 14120.5 & 77.5\\ 
		7 & 14127.2 & 74.8 & 14164.1 & 75.1 & 14198.0 & 75.2\\ 
		8 & 14202.0 & 71.2 & 14239.2 & 71.6 & 14273.2 & 71.5\\ 
		9 & 14273.2 & 71.5 & 14310.8 & 68.7 & 14344.7 & 68.0\\ 
		10 & 14344.7 & 66.3 & 14379.5 & 64.8 & 14412.5 & 64.1\\ 
		11 & 14411.0 & 64.2 & 14444.3 & 65.2 & 14476.6 & 63.0\\ 
		12 & 14475.2 & 62.1 & 14509.5 & 59.4 & 14539.6 & 58.8\\ 
		13 & 14537.3 & 59.4 & 14568.9 & 56.9 & 14598.4 & 55.8\\ 
		14 & 14596.7 & 55.1 & 14625.8 & 54.7 & 14654.2 & 51.4\\ 
		15 & 14651.8 & 51.4 & 14680.5 & 51.4 & 14705.6 & 46.2\\ 
		16 & 14703.2 & 47.6 & 14731.9 & 44.3 & 14751.4 & 43.4\\ 
		17 & 14750.8 & 47.0 & 14776.2 & 40.1 & 14795.2 & 37.9\\ 
		18 & 14797.8 & 45.9 & 14816.3 & 35.5 & 14833.1 & 32.5\\ 
		19 & 14843.7 & 25.2 & 14851.8 & 24.4 & 14865.6 & 23.9\\ 
		20 & 14868.9 & 21.3 & 14876.2 & 16.6 &14889.5 & 8.7\\ 
		21 & 14890.2 & 10.7 & 14892.8 & 6.9 & 14898.2 & \\ 
		22 & 14900.9 & & 14899.7 & &         &    \\ \hline \hline
	\end{tabular}
	\caption{Experimental assignments for the rotationless energy of the vibrational levels of the $d \ ^3\Pi$ state based on our RE2PI data, aided by our depletion data for $v^{\prime}=0$ and 1.  Uncertainty for all assignments is 0.5 cm$^{-1}$.  We have referenced the term energies, $T_v$, to the Rb 5S + Li 2S asymptote.}
	\label{tab:masterAssignments}
\end{table*}
\section{Discussion}
\begin{figure} [b]
	\includegraphics[width=8.6cm]{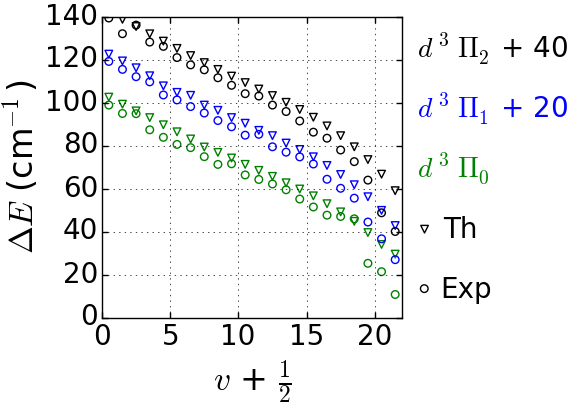}\\
	\caption{(Color on-line) Comparison of our extracted vibrational splitting to predicted vibrational splitting.  The circles represent our data, while the triangles are predicted by \textit{ab initio} curves~\cite{Korek}.  Green markers label $\Omega=0$ (compared to \textit{ab initio} $\Omega=0^+$) spacings.  Blue markers are shifted by +20 cm$^{-1}$ and label $\Omega=1$ spacings.  Black markers are shifted by +40 cm$^{-1}$ and label $\Omega=2$ spacings.}
	\label{fig:vibrationalSplitting}
\end{figure}

We determine the vibrational binding energies and rotational constants of the states seen in depletion spectroscopy, which we tabulate in Table.~\ref{tab:depletionAssignments}.  We list in Table~\ref{tab:masterAssignments} the  assignments and energy of each of the $d \ ^3\Pi$ states that we observe through RE2PI and depletion spectroscopy.  We also include in this table the energy difference between adjacent states, which aids in the assignment of the lines.  

The theoretical vibrational levels and spin-orbit splittings that we used to guide our work and for comparison of results come from \textit{ab initio} calculations by Korek et al.~\cite{Korek} with aid from LEVEL 8.0~\cite{LEVEL}.  We found good overall agreement with these \textit{ab initio} results.  The spin-orbit splittings for $\Omega=0$ to $\Omega=1$, predicted to be 21 cm$^{-1}$, are measured here to be 37 cm$^{-1}$; for $\Omega=1$ to $\Omega=2$, they are predicted to be 38 cm$^{-1}$, and we found them to be 33 cm$^{-1}$.  For the spin-orbit splitting between the $\Omega=0^{+}$ to $\Omega=0^{-}$ states, however, we observe 0.9 cm$^{-1}$, significantly less than the predicted 36 cm$^{-1}$.  Our depletion data is unambiguous in establishing the $\Omega=0^{+}$ to $\Omega=0^{-}$ splitting, and a small $\Omega=0^{+}$ to $\Omega=0^{-}$ splitting is consistent with our observations in the $(3) \ ^3\Pi$ state~\cite{Adeel}.

\begin{table*}[t]
	\centering
	\begin{tabular}{ccccccc}
		\hline \hline
		\multirow{2}{*}{} & \multicolumn{2}{c}{$d \ ^3\Pi_0$} & \multicolumn{2}{c}{$d \ ^3\Pi_1$} & \multicolumn{2}{c}{$d \ ^3\Pi_2$} \\
		& Exp. & Th. & Exp. & Th. & Exp. & Th. \\ \midrule[1.5pt]
		$T_e$ (cm$^{-1}$) & 13459.2 (1.5) & 13300.7 & 13497.5 (2.0) & 13359.9 &  13528.0 (1.4) & 13398.2\\ 
		$\omega_e$ (cm$^{-1}$) & 101.4 (0.7) & 103.6 & 100.4 (0.9) & 102.8 & 101.7 (0.6) & 102.4\\ 
		$x_e$ ($10^{-3}$) & 16.7 (0.8) & 13.1 & 15.0 (1.0) & 12.7 & 15.8 (0.7) & 12.5\\ 
		$y_e$ ($10^{-3}$) & 0.068 (0.028) & -0.06 & -0.038 (0.033) & -0.10 & -0.036 (0.026) & -0.12\\ \hline \hline
	\end{tabular}
	\caption{Molecular vibrational constants fitted to our data, $T(v) = T_e + \omega_e (v + 1/2) - \omega_e x_e (v + 1/2)^2 + \omega_e y_e (v + 1/2)^3$ where $T(v)$ is the rotationless energy of the $v^{th}$ vibrational level.  The uncertainty is given in parentheses.  The theory values are from fitting the bound states calculated by LEVEL 8.0 using PECs from Ref.~\cite{Korek}.  When fitting the experimental data, we used only the $v^{\prime}$=0-19 to increase the accuracy.}
	\label{tab:molecularConstants}
\end{table*}

We show the vibrational spacing, $\Delta E = E_{v+1} - E_v$ vs $v$, of the different series in Fig.~\ref{fig:vibrationalSplitting}.  These data are in reasonable agreement with the predicted vibrational splittings although there appears to be a nearly uniform difference of a few cm$^{-1}$.  We found that the depth of the $d \ ^3\Pi$ potential (exp. value) is less than that predicted (th. value).  We looked extensively for another vibrational level below our assigned $v^{\prime}=0$ level.  We covered $\pm$ 10 cm$^{-1}$ around the expected vibrational location with our depletion spectra, but found no indication of a depletion resonance. 

Our extracted molecular constants are listed in Tab.~\ref{tab:molecularConstants}.  These provide an easy estimation of the spectral structure of the $d \ ^3\Pi$ states as well as a quick comparison to theoretical predictions.  As borne out in Fig.~\ref{fig:vibrationalSplitting}, there is good agreement between our fitted harmonic constant, $\omega_e$, and the predictions.  However, there is considerably less agreement between our extracted term energy, $T_e$, and the predictions for reasons discussed previously.  Additionally, it is important to note that when we fitted the experimental data to determine $T_e$, $\omega_e$, $x_e$ and $y_e$ we used only $v^{\prime}$=0-19.  This increased the accuracy of the fit so that for these vibrational levels our molecular constants reproduce our data with a standard deviation of 2 cm$^{-1}$.  We believe most of the deviations are caused by experimental uncertainties on determining the frequencies of the peaks, as well as small perturbations to state locations caused by spin-orbit mixing.

\section{Conclusion}

In this work we have found and identified the $v=0$ to $v=22$ states of the $d \ ^3\Pi_{\Omega}$ electronic state in LiRb.  We explored singlet - triplet mixing to evaluate the possibility of using the $d \ ^3\Pi$ - $D \ ^1\Pi$ mixed states as the intermediate state in a STIRAP transfer from a weakly bound triplet state to deeply bound singlet states.  We know from heat pipe spectra~\cite{Ivanova11} and other REMPI data~\cite{Lorenz14} that the transitions from deeply bound $D \ ^1\Pi \ v^{\prime}$ to deeply bound $X \ ^1\Sigma^+$ states are strong.  Our depletion data shows that transitions from weakly bound triplet molecules to deeply bound $d \ ^3\Pi_1 \ v^\prime$ states are also strong.  Finally, our RE2PI data demonstrate that there is about 10\% mixing between most of these singlet and triplet states.  We suggest using the $d \ ^3\Pi_1 \ v^\prime = 0 \ J^\prime = 1$ state as the intermediate state for a STIRAP transfer in future work.  The laser wavelengths for this transfer would be 740 nm, achievable with Ti:Sapphire or diode lasers, and 516 nm, which is accessible with green laser diodes.  Using calculated FCFs and an estimate of the transition dipole moment of a few times $e a_0$, the `up' transition would be around $4 \times 10^{-2}$ $e a_0$ and the `down' transition would be around $10^{-2}$ $e a_0$ which is competitive with the transfer strength used by the KRb JILA team~\cite{Ni08}.

Using the binding energy of the lowest several vibrational levels we have assessed their potential for short-range photoassociation.  There are only six possible PA transitions we can observe with our current Ti:Sapphire laser, the same six we saw in depletion.  However, our depletion data provides the perfect stepping stone for this type of work, providing very trusted locations of vibrational levels (to within 0.5 cm$^{-1}$, the uncertainty in the binding energy of $a \ ^3\Sigma^+ \ v^{\prime \prime}=13$).  Of particular interest to us is the combination of short range PA and mixing between $d \ ^3\Pi$ - $D \ ^1\Pi$.  This provides an avenue through simple spontaneous decay to the rovibronic ground state although we will need a different laser to access these PA resonances.

We are happy to acknowledge useful conversations with Jes\'{u}s P\'{e}rez-R\'{\i}os, and university support of this work through the Purdue OVPR AMO incentive grant.  And we would like to acknowledge the work done by S. Dutta and J. Lorenz in building the LiRb machine.
\bibliography{Literature}
\end{document}